\documentstyle[aps,preprint]{revtex}

\begin{document}
\title{Wigner lattice of ripplopolarons in a multielectron bubble in helium}
\author{J. Tempere$^{1,2}$, S. N. Klimin$^1$,
 I. F. Silvera$^2$, J. T. Devreese$^{1,\ast}$}
\address{$^1$ TFVS, Universiteit Antwerpen, Universiteitsplein 1, B2610 Antwerpen, Belgium.}
\address{$^2$ Lyman Laboratory of Physics, Harvard University, Cambridge MA02138, USA.}
\address{author e-mail addresses: jtempere@uia.ua.ac.be, klimin@uia.ua.ac.be,
silvera@physics.harvard.edu, devreese@uia.ua.ac.be}
\date{Appeared in European Physical Journal B \textbf{32}, 329 (2003).}
\maketitle

\begin{abstract}
The properties of ripplonic polarons in a multielectron bubble in 
liquid helium are investigated on the basis of a path-integral 
variational method. We find that the two-dimensional electron gas can form 
deep dimples in the helium surface, or ripplopolarons, to solidify as a 
Wigner crystal. We derive the experimental conditions of temperature, 
pressure and number of electrons in the bubble for this phase to be 
realized. This predicted state is distinct from the usual Wigner lattice 
of electrons, in that it melts by the dissociation of the ripplopolarons, 
when the electrons shed their localizing dimple as the pressure on the 
multielectron bubble drops below a critical value.
\end{abstract}


\tighten

\section{Introduction}
The two-dimensional (2D) electron system formed on the surface of liquid 
helium has been widely investigated, especially with regard to the 
formation and melting of a Wigner lattice \cite{AndreiBook}. 
An electric field pressing the electrons against the helium surface 
results in an interaction between the electrons and the quantized modes 
of oscillation of the helium surface, the ripplons \cite{ShikinJETP31}. 
In this paper, we investigate the effects of the electron-ripplon 
coupling in a multielectron bubble in liquid helium and highlight the 
differences with the case of electrons on a flat helium surface. Such 
multielectron bubbles (MEBs) are spherical cavities in liquid helium. 
MEBs are typically 0.1 $\mu $m -- 100 $\mu $m in radius for 
10$^{3}-$10$^{8}$ electrons \cite {VolodinJETP26}. The bubble's diameter 
is determined by the balance between the surface tension of liquid helium 
and the Coulomb repulsion of the electrons \cite{ShikinJETP27}. 

The electrons are not distributed throughout the volume of the bubble, but 
form a nanometer thin and effectively 2D layer anchored to 
the inside of the bubble surface \cite{SalomaaPRL47}. Although the flat
2D electron gas has been studied extensively and new  
quantum states such as the fractional quantum Hall regime were 
revealed, much less effort has gone to the study of the spherically 
curved 2D electron gas such as the one present in MEBs. 
Spherical shells of charged particles appear also in a variety of other 
physical systems, such as fullerenes \cite{RotkinSSSP36}, metallic 
nanoshells \cite{AverittPRL78}, and charged macroscopic droplets. The 
main goal of this paper is to show that the enhanced electron-ripplon 
coupling in the bubble leads to a new solid phase, a lattice of 
ripplonic polarons, that is distinct from the electron Wigner lattice, 
and to investigate the properties of this phase.

Using pressure, the surface of the MEB can be compressed to achieve 2D
electron densities as high as $10^{14}$ cm$^{-2}$ \cite{TemperePRL87},
whereas flat surfaces are limited to $~ 2 \times 10^9$ cm$^{-2}$ due to an
instability \cite{GorkovJETP18}. This instability is not present in
multielectron bubbles \cite{TemperePRB67}. As a result very large electric
fields exist on an electron, normal to the surface (due to all other
electrons in the MEB), whereas for a flat surface the maximum field is
around 3 kV/cm.  Some typical values for physical variables related to the
bubble are given in table \ref{tab:1}.

The Hamiltonian of a single electron on a flat helium surface is given 
by 
\begin{equation}
  \hat{H}= \frac{\hat{p}^{2}}{2m_{e}}
  +\sum_{q}\omega (q)\hat{a}_{{\bf q}}^{+}\hat{a}_{{\bf q}}+\sum_{q}M_{q}
  e^{-i {\bf q.r}}\left( \hat{a}_{{\bf q}}+\hat{a}_{-{\bf q}}^{+}\right),
\label{H0}
\end{equation}
where ${\bf \hat{p}}$ is the electron momentum operator, 
$m_{e}$ is the electron mass, and 
$\omega (q)=q\sqrt{\sigma q/\rho }$ is the ripplon dispersion relation 
with $\sigma \approx 3.6 \times 10^{-4}$ J/m$^{2}$ the surface tension 
of helium and $\rho =145$ kg/m$^{3}$ the mass density of helium. In 
this Hamiltonian, we restrict ourselves to 2D position and 
momentum operators, assuming that the part of the wave function of the 
electrons relating to the direction perpendicular to the surface can
be factored out exactly. The second-quantization operators 
$\hat{a}_{{\bf q}}^{+},\hat{a}_{{\bf q}}$ create/annihilate a ripplon 
with planar wave number ${\bf q}$. The electron-ripplon coupling 
amplitude is given by 
\begin{equation}
  M_{q}=\sqrt{ \frac{\hbar q}{2\rho \omega (q)}} e|{\bf E}|,
\end{equation}
where ${\bf E}$ is the electric field perpendicular to the surface (the
so-called `pressing field'), and $e$ is the electron charge. The pressing
field pushes the electrons with a force $e{\bf E}$ towards the helium
surface, that acts like a sheet with surface tension $\sigma $. Note that
there is a 1 eV barrier preventing single electrons from penetrating the
helium surface. The self-induced trapping potential of the electron on the
helium surface is manifested by the appearance of a dimple in the helium
surface underneath the electron, much like the deformation of a rubber
sheet when a person is pulled down on it by a gravitational force. The
resulting quasiparticle consists of the electron together with its dimple
and can be called a ripplonic polaron or ripplopolaron
\cite{ShikinJETP38}.

Hamiltonian (\ref{H0}) for the ripplopolarons is very similar to the
Fr\"{o}hlich Hamiltonian describing polarons \cite{FrohlichAdvP3}; the
role of the phonons is now played by the ripplons. Methods suitable for
the study of single polarons have been used to analyse the single
ripplopolaron on a flat surface \cite{JacksonPRB24,JacksonPRB30}.
Recently, Fratini and Qu\'{e}merais \cite{FratiniEPJB14} have proposed a
path integral treatment for a Wigner lattice of polarons. One of the goals
of the present paper is to adapt their method so that it becomes suitable
for the treatment of a lattice of ripplopolarons.

In Sec. \ref{sec:1} we introduce the Hamiltonian describing a
ripplopolaron in a Wigner lattice of ripplopolarons in a multielectron
bubble. The temperature zero solution for the ground state of this
Hamiltonian is derived in the strong-coupling case in Sec. \ref{sec:2}.
The results for arbitrary interaction strength and arbitrary finite
temperature, obtained with the variational path-integral technique, are
presented in Sec. \ref{sec:3}. These finite-temperature results allow us
to investigate the melting of the ripplopolaron Wigner lattice and
determine the phase diagram of this state in Sec. \ref{sec:4}.  The
results are discussed and compared with the case of electrons on a flat
helium surface in Sec. \ref{sec:5}.

\section{Hamiltonian for a ripplopolaron in a Wigner lattice}
\label{sec:1}

In their treatment of the electron Wigner lattice embedded in a polarizable
medium such as a semiconductors or an ionic solid, Fratini and Qu\'{e}merais 
\cite{FratiniEPJB14} described the effect of the electrons on a particular
electron through a mean-field lattice potential. The (classical) lattice
potential $V_{lat}$ is obtained by approximating all the electrons
acting on one particular electron by a homogenous charge density in which a
hole is punched out; this hole is centered in the lattice point of the
particular electron under investigation and has a radius given by the
lattice distance $d$. Thus, in their approach, the anisotropy effects are
neglected. A second assumption implicit in this approach is that the effects
of exchange are neglected. This can be justified by noting that for the
electrons to form a Wigner lattice it is required that their wave 
function be localized to within a fraction of the lattice parameter as 
follows from the Lindemann criterion \cite{LindemanZPhys11}. As can be 
read from table \ref{tab:1}, the typical distance between electrons (the 
lattice parameter) is 10-100 nm.

Within this particular mean-field approximation, the lattice potential can
be calculated from classical electrostatics and we find that for a
2D electron gas it can be expressed in terms of the elliptic
functions of first and second kind, $E\left( x\right) $ and $K\left(
x\right)$,
\begin{eqnarray}
  V_{lat}\left({\bf r}\right) &=&-\frac{2e^{2}}{\pi d^{2}}\left\{ \left|
  d-r\right| E\left[ -\frac{4rd}{\left( d-r\right) ^{2}}\right]  
  \right. 
\nonumber \\
  && \left. +\left(d+r\right) \mathop{\rm sgn} \left( d-r\right) 
  K\left[ -\frac{4rd}{\left( d-r\right) ^{2}}\right] \right\} .  
\label{Potential}
\end{eqnarray}
Here, {\bf r} is the position vector measured from the lattice position. We
can expand this potential around the origin to find the small-amplitude
oscillation frequency of the electron lattice: 
\begin{equation}
  \lim_{r\ll d}V_{lat}\left( {\bf r}\right) =-\frac{2e^{2}}{d}+\frac{1}
  {2}m_{e}\omega _{lat}^{2}r^{2}+{\cal O}\left( r^{4}\right) ,
\label{Potlimit}
\end{equation}
with the confinement frequency 
\begin{equation}
  \omega _{lat}=\sqrt{\frac{e^{2}}{m_{e}d^{3}}}.
\label{phonfreq}
\end{equation}
Although this mean-field approach may seem crude, it has the distinct
advantage that the `phonon' frequency $\omega _{lat}$ of the electron
lattice corresponds closely to the longitudinal plasmon frequency that can
be derived using an entirely different approach based on a more rigorous
study of the modes of oscillations of both the bubble and the charge
distribution on the bubble surface. This frequency lies typically in the THz
range and the lattice parameter $d$ in MEBs ranges roughly from 10 to 
100 nm. From this, and from the succesful application of this mean-field
approach to polaron crystals in solids, we conclude that the approach based
on that of Fratini and Qu\'{e}merais describes the influence of the other
electrons well in the framework of small amplitude oscillations of the
electrons around their lattice point. The (modified) Lindemann melting
criterion suggests that the lattice will melt when the electrons are on
average displaced more than ca. 10\% from their lattice position; thus in
the regime of interest the Fratini-Qu\'{e}merais approach is applicable. 
In the mean-field approximation, the Hamiltonian for a ripplopolaron in a 
lattice on a \textit{locally flat} helium surface is given by
\begin{eqnarray}
  \hat{H}&=& {\displaystyle{\hat{p}^{2} \over 2m_{e}}}
  +V_{lat}\left( {\bf \hat{r}}\right) +\sum_{{\bf q}}\hbar \omega (q)
  \hat{a}_{{\bf q}}^{+}\hat{a}_{{\bf q}} 
\nonumber \\
  &&+\sum_{{\bf q}}M_{q}e^{-i{\bf q.r}}\left( \hat{a}_{{\bf q}}
  +\hat{a}_{-{\bf q}}^{+}\right) ,  
\label{H1}
\end{eqnarray}
where ${\bf \hat{r}}$ is the electron position operator.

Now that the lattice potential has been introduced, we can move on and
include effects of the bubble geometry. If we restrict our treatment to the
case of large bubbles (with $N>10^{5}$ electrons) such as those already
experimentally observed \cite{VolodinJETP26}, then both the ripplopolaron
radius and the inter-electron distance $d$ are much smaller than the radius
of the bubble $R_{b}$. This gives us ground to use the locally flat
approximation using the auxiliary model of a ripplonic polaron in a planar
system described by (\ref{H1}), but with a modified ripplon dispersion
relation and an modified pressing field. In appendix A we provide a more
detailed description of the electron-ripplon interaction on a sphere and
discuss how the locally flat approximation (\ref{H1}) is linked to the exact
expressions for the curved surface. Essentially, we find for the modified
ripplon dispersion relation in the MEB: 
\begin{equation}
  \omega (q)=\sqrt{{\displaystyle{\sigma  \over \rho }}
  q^{3}+{\displaystyle{p \over \rho R_{b}}} q}, 
\label{ripplodisp}
\end{equation}
where $R_{b}$ is the equilibrium bubble radius which depends on the
pressure and the number of electrons \cite{TemperePRL87}. The bubble radius
is found by balancing the surface tension and the pressure with the Coulomb
repulsion \cite{TemperePRL87}. At zero pressure, it scales as $N^{2/3}$ and
in the pressure-dominated regime ($p\gg \sigma /R_{b}$) it scales with
pressure as $p^{-1/4}$. Some typical values are given in table I. The
modified electron-ripplon interaction amplitude in an MEB is given by
\begin{equation}
  M_{{\bf q}}=e|{\bf E}|\sqrt{ {\displaystyle{\hbar q \over 
  2\rho \omega (q)}} }.  
\label{elripcoupl}
\end{equation}
The effective electric pressing field pushing the electrons against the
helium surface and determining the strength of the electron-ripplon
interaction is 
\begin{equation}
  {\bf E}=-{\displaystyle{Ne \over 2R_{b}^{2}}} {\bf e}_{{\bf r}}. 
\label{pressfield}
\end{equation}
Some typical values for the pressing field and the dimensionless coupling
constants are given in Table I. Note that the electric field in the bubble
is larger than the typical pressing fields (of the order of $10^{2}-10^{3}$
V/cm) applied on electrons on a flat helium surface. Thus, the
electron-ripplon coupling will be stronger in the multielectron bubble. The
modified ripplon dispersion relation (and the dependence of the bubble
radius on the number of electrons and the pressure) was studied in more
detail in Ref. \cite{TemperePRL87}, and the stability of the multielectron
bubble against surface deformational modes was investigated in detail in
Refs. \cite{SalomaaPRL47,TemperePRB67}. These studies concluded that even
though a large effective electric pressing field is present at zero
pressure, the bubbles can be stable in contrast to the flat surface which
can only sustain a moderate electric pressing field.

The crucial differences that exist between the case of a ripplopolaron in
the multielectron bubble and on the flat surface (and that are preserved in
the locally flat approximation) are (i) the electric pressing field 
${\bf E}$ is stronger than that typically realised for electrons on 
helium films (see table I) and thus the electron-ripplon coupling is
enhanced as compared to the normal film; (ii) the interaction energy
arising from the change in polarisation of the helium due to the
displacement of the electron has a similar form, but is much weaker and 
can be neglected. In addition the electric field, and thus the 
electron-ripplon coupling increases as the bubble radius is decreased. 
Thus, pressurizing the bubbles, which decreases the radius, also increases 
the electron-ripplon coupling strength (rougly as $R_{b}^{-2}$). In the
high-pressure regime ($p\gg \sigma /R_{b}$), the bubble radius goes as
$p^{-1/4}$ and thus the electron-ripplon coupling increases as $\sqrt{p}$.
The pressure provides a `tuning knob' to set the
electron-ripplon interaction strength at a desired level.

\section{The Ripplopolaron Wigner Lattice at temperature zero}
\label{sec:2}

\subsection{Ground state}

To gain insight into the nature of the Wigner solid of ripplopolarons, we
will analyse Hamiltonian (\ref{H1}) first in the strong-coupling approach.
In the next section, the more general and more accurate Feynman
variational pa\-th-in\-te\-gral method will be applied, generalizing the
results of this section to finite temperature. Noting that the 
frequency associated with the electron's motion, $\omega _{lat}$, is
typically several orders of magniture larger than the frequency associated
with the ripplons \cite{KliminSSC}, $\omega (q)$, we can safely make the
product ansatz for the wave function of the ripplopolaron in the lattice:
$\left| \Psi \right\rangle =\left| \psi _{e}\right\rangle \left| \phi
\right\rangle $. Here $\left| \phi \right\rangle $ is the factor of the
wave function that contains the ripplon coordinates, and $\left| \psi
_{e}\right\rangle $ is the electronic part of the wave function. For
small-amplitude oscillations of the electrons around their lattice site,
the lattice potential $V_{lat}$ is well approximated by a parabolic
potential (\ref{Potlimit}), so we choose a Gaussian trial wave function
for the electronic part:
\begin{equation}
  \left| \psi_{e}\right\rangle = {\displaystyle{1 \over \pi ^{1/2}a}}
  e^{-r^{2}/(2a^{2})}.
\end{equation}
In this trial wave function, the variational parameter is $a$, the width of
the electron wave function. Taking the expectation value of Hamiltonian (
\ref{H1}) with respect to this electronic part of the wave function yields: 
\begin{eqnarray}
  \left\langle \psi _{e}\right| \hat{H}\left| \psi _{e}\right\rangle  &=&
  {\displaystyle{\hbar ^{2} \over 2m_{e}a^{2}}}
  +\frac{m_{e}\omega _{lat}^{2}}{2}a^{2}  
  +\sum_{{\bf q}}\hbar \omega (q)\hat{a}_{{\bf q}}^{+}\hat{a}_{{\bf q}}
\nonumber \\
  && +\sum_{{\bf q}}M_{q}e^{-a^{2}q^{2}/4}\left( \hat{a}_{{\bf q}}
  +\hat{a}_{-{\bf q}}^{+}\right) ,
\end{eqnarray}
The ripplonic part of $\left\langle \psi _{e}\right| \hat{H}\left| \psi
_{e}\right\rangle $ represents a displaced harmonic oscillator and can be 
rewritten as
\begin{eqnarray}
 \lefteqn{
  \left\langle \psi _{e}\right| \hat{H}\left| \psi _{e}\right\rangle  =
  {\displaystyle{\hbar ^{2} \over 2m_{e}a^{2}}}
  +\frac{m_{e}\omega _{lat}^{2}}{2}a^{2}-\sum_{{\bf q}}
  {\displaystyle{|M_{q}|^{2}e^{-a^{2}q^{2}/2} \over \hbar \omega (q)}}
 } & & 
\nonumber \\
 & & +
 \sum_{{\bf q}}\hbar \omega (q)\left[ \hat{a}_{{\bf q}}^{+}+
 {\textstyle{M_{q}e^{-a^{2}q^{2}/4} \over \hbar \omega (q)}} \right] 
 \left[ \hat{a}_{{\bf q}} +
 {\textstyle{M_{q}e^{-a^{2}q^{2}/4} \over \hbar \omega (q)}} \right] . 
\label{pekar3}
\end{eqnarray}
The ground state of the displaced (2D) harmonic oscillator at temperature
zero has energy $\hbar \omega (q)$, independent of the variational parameter 
$a$. To find the variational optimal value of $a$, we minimize the
ripplopolaron energy 
\begin{equation}
  E(a)={\displaystyle{\hbar ^{2} \over 2m_{e}a^{2}}}
  +\frac{m_{e}\omega _{lat}^{2}}{2}a^{2}-\sum_{{\bf q}}
  {\displaystyle{|M_{q}|^{2}e^{-a^{2}q^{2}/2} \over \hbar \omega (q)}}.  
\label{Epekar}
\end{equation}
The sum over momenta can be rewritten as an integral, remembering 
\begin{equation}
  \sum_{{\bf q}}\rightarrow 
  \displaystyle\int \limits_{q>1/R_{b}}
  {\displaystyle{d^{2}{\bf q} \over (2\pi )^{2}}}
  = \displaystyle\int \limits_{\tilde{q}>1}
{\displaystyle{d^{2}{\bf \tilde{q}} \over (2\pi )^{2}R_{b}^{2}}}.
\end{equation}
The lower limit appears since the largest wavelength available is $1/R_{b}$.
We checked that the final results do not depend crucially on the value of
this naturally occurring cut-off. A dimensionless integration variable ${\bf 
\tilde{q}}={\bf q}R_{b}$ is introduced. The ground state energy of the
ripplopolaron can then be evaluated analytically: 
\begin{eqnarray}
E(a) &=& {\displaystyle{\hbar ^{2} \over 2m_{e}a^{2}}}
         +\frac{m_{e}\omega _{lat}^{2}}{2}a^{2}  \\
 & & -{\displaystyle{(e|{\bf E}|)^{2} \over 2\pi \sigma }}
     \displaystyle\int \limits_{1}^{\infty }d\tilde{q}
     {\displaystyle{\tilde{q} \over \tilde{q}^{2}+
     {\displaystyle{pR_{b} \over \sigma }}}}
     \exp \left[ -{\displaystyle{a^{2} \over 2R_{b}^{2}}}
     \tilde{q}^{2}\right] \nonumber \\
 &=& {\displaystyle{\hbar ^{2} \over 2m_{e}a^{2}}}
     +\frac{m_{e}\omega _{lat}^{2}}{2}a^{2} \label{Eripol} \\
 & & -{\displaystyle{(e|{\bf E}|)^{2} \over 4\pi \sigma }}
     \exp \left[ {\displaystyle{pa^{2} \over 2\sigma R_{b}}}
     \right] \Gamma \left[ 0,
     {\displaystyle{a^{2} \over 2R_{b}^{2}}}
     \left( 1+{\displaystyle{pR_{b} \over \sigma }}
     \right) \right] , \nonumber 
\end{eqnarray}
where $\Gamma $ is the incomplete gamma function. Fig. \ref{fig:1} shows
the result for the variational parameter $a$ as a function of number of
electrons and pressure in the multielectron bubble. If this figure, $a$ is
expressed relative to the interelectron distance $\sqrt{4\pi
R_{b}^{2}/N}$.

\subsection{Dimple shape}

The ripplonic part of the Hamiltonian (\ref{pekar3}) represents oscillations
of the helium surface no longer around the original bubble surface, but
around a new, displaced equilibrium surface. This displacement of the helium
surface is the dimpling. Underneath each electron, a dimple appears. The new
equilibrium surface, described by the function $u({\bf r})$ (cf. 
appendix, Eq. \ref{radius}), can be found by using the canonical relation 
between the surface displacement operator and the ripplon creation and 
annihilation operators: 
\begin{equation}
  \hat{Q}_{{\bf q}}=\sqrt{{\displaystyle{\hbar q \over 2\rho\omega(q)}}} 
  (\hat{a}_{{\bf q}}+\hat{a}_{-{\bf q}}^{+}),
\end{equation}
and evaluating 
\begin{equation}
u({\bf r})=\sum_{{\bf q}}\left\langle \Psi \right| \hat{Q}_{{\bf q}}\left|
\Psi \right\rangle e^{i{\bf q}.{\bf r}}.
\end{equation}
The result is given by 
\begin{equation}
  u({\bf r})= {\displaystyle{e|{\bf E}| \over 2\pi \sigma }}
  \int_{1}^{\infty }d\tilde{q} {\displaystyle{\tilde{q} \over 
  \tilde{q}^{2}+{\displaystyle{pR_{b} \over \sigma }}}}
  J_{0}\left( {\displaystyle{\tilde{q}r \over R_{b}}}
  \right) e^{-a^{2}\tilde{q}^{2}/(4R_{b}^{2})}.  
\end{equation}
In the limiting case of a large bubble, this result corresponds to that of
Shikin and Monarkha \cite{ShikinJETP38} for electrons on a flat helium
surface; the role of the capillary constant is played by $p/(\sigma
R_{b})$. Fig. \ref{fig:2} shows, for a bubble with $N=10^5$ electrons, at
different pressures the shape of the dimpled surface. Several dimples are
shown -- above the center of each dimple an electron is present. The
dotted curve represents the undimpled $u({\bf r})=0$ surface; the
curvature of the bubble surface is visible in this curve. The electrons
are separated by the interelectron distance $d=\sqrt{4\pi R_{b}^{2}/N}$.
As the pressure increases, the radius of the bubble decreases. Since the
number of electrons is fixed the electric pressing field increases, making
on its turn the electron-ripplon coupling larger. This results in deeper,
narrower dimples. Note that while the deformation here can be several
angstroms, for a flat surface on bulk helium the maximum deformation of a
dimple is less than one angstrom \cite{IkeziPRB25}. Also for electrons on
a thin helium film above a dielectric substrate, the dimple depth can
reach several angstrom \cite{MarquesPRB39}.

\section{The Ripplopolaron Wigner Lattice at finite temperature}
\label{sec:3}

The simple but intuitive approach of the previous section describes the
system in the limit of temperature zero. To study the ripplopolaron Wigner
lattice at finite temperature (and for any value of the electron-ripplon
coupling), we use the variational path-integral approach \cite{Feynman}.
This variational principle distinguishes itself from Rayleigh-Ritz
variation in that it uses a trial action functional $S_{trial}$ instead
of a trial wave function.

The action functional of the system described by Hamiltonian (\ref{H1}), 
becomes, after elimination of the ripplon degrees of freedom,
\begin{eqnarray}
S &=&- {\displaystyle{1 \over \hbar }}
       \displaystyle\int \limits_{0}^{\hbar \beta }d\tau \left\{ 
       {\displaystyle{m_{e} \over 2}}
       \dot{r}^{2}(\tau )+V_{lat}[r(\tau )]\right\}  
   +\sum_{{\bf q}}\left| M_{q}\right| ^{2}
\nonumber \\
&& \times
\displaystyle\int \limits_{0}^{\hbar \beta }d\tau 
\displaystyle\int \limits_{0}^{\hbar \beta }d\sigma 
G_{\omega (q)}(\tau -\sigma ) 
e^{ i{\bf q}\cdot \lbrack {\bf r}(\tau )-{\bf r}(\sigma )] },  \label{S}
\end{eqnarray}
with 
\begin{equation}
  G_{\nu }(\tau -\sigma )={\displaystyle{\cosh [\nu (|\tau -\sigma |-\hbar 
  \beta /2)] \over \sinh (\beta \hbar \nu /2)}}.
\end{equation}
In preparation of its customary use in the Jensen-Feynman inequality, the
action functional (\ref{S}) is written in imaginary time $t=i\tau $ with $%
\beta =1/(k_{B}T)$ where $T$\thinspace is the temperature. Following an
approach analogous to that of Fratini and Qu\'{e}merais for a lattice of
polarons in a semiconductor \cite{FratiniEPJB14}, and to that of Devreese 
et al. for $N$ polarons in a quantum dot \cite{DevreeseSSC114}, we 
introduce a quadratic trial action of the form 
\begin{eqnarray}
  S_{trial}&=&-{\displaystyle{1 \over \hbar }} \displaystyle\int 
   \limits_{0}^{\hbar \beta }d\tau \left[ {\displaystyle{m_{e} \over 2}}
   \dot{r}^{2}(\tau ) + {\displaystyle{m_{e}\Omega ^{2} \over 2}}
   r^{2}(\tau )\right] 
\nonumber \\
  && - {\displaystyle{Mw^{2} \over 4\hbar }} \displaystyle\int 
   \limits_{0}^{\hbar \beta }d\tau \displaystyle\int 
   \limits_{0}^{\hbar \beta }d\sigma 
   G_{w}(\tau -\sigma ){\bf r}(\tau )\cdot 
   {\bf r}(\sigma ).  
\label{S0}
\end{eqnarray}
where $M,w,$ and $\Omega $ are the variationally adjustable parameters. This
trial action corresponds to the Lagrangian 
\begin{equation}
  {\cal L}_{0}= {\displaystyle{m_{e} \over 2}} \dot{r}^{2}
  +{\displaystyle{M \over 2}} \dot{R}^{2}
  -{\displaystyle{\kappa  \over 2}} r^{2}
  -{\displaystyle{K \over 2}} ({\bf r}-{\bf R})^{2},  
\label{L0}
\end{equation}
from which the degrees of freedom associated with ${\bf R}$ have been
integrated out. This Lagrangian can be interpreted as describing an 
electron with mass $m_{e}$ at position ${\bf r}$, coupled through a spring 
with spring constant $\kappa$ to its lattice site, and to which a 
fictitious mass $M$ at position ${\bf R}$ has been attached with another 
spring, with spring constant $K$. The relation between the spring 
constants in (\ref{L0}) and the variational parameters $w,\Omega $ is 
given by 
\begin{eqnarray}
  w &=&\sqrt{K/m_{e}}, 
\\
  \Omega &=&\sqrt{(\kappa +K)/m_{e}}.
\end{eqnarray}

Based on the trial action $S_{trial}$, Feynman's variational method
allows one to obtain an upper bound for the free energy $F$ of the system
(at temperature $T$) described by the action functional $S$ by minimizing 
the following function:

\begin{equation}
  F=F_{0}-\frac{1}{\beta} \left\langle S-S_{trial}\right\rangle,  
\label{JF}
\end{equation}
with respect to the variational parameters of the trial action. In this
expression, $F_{0}$ is the free energy of the trial system characterized by
the Lagrangian ${\cal L}_{0}$, $\beta=1/(k_b T)$ is the inverse 
temperature, and the expectation value $\left\langle S-S_{%
trial}\right\rangle$ is to be taken with respect to the ground state
of this trial system. The evaluation of expression (\ref{JF}) is
straightforward though lengthy. We find 
\begin{eqnarray}
\lefteqn{ F =
   {\displaystyle{2 \over \beta }} \ln \left[ 2\sinh \left( 
   {\displaystyle{\beta \hbar \Omega _{1} \over 2}}\right) \right] 
 + {\displaystyle{2 \over \beta }} \ln \left[ 2\sinh \left( 
   {\displaystyle{\beta \hbar \Omega _{2} \over 2}}\right) \right] 
} && \nonumber \\
&&-{\displaystyle{2 \over \beta }} \ln \left[ 2\sinh \left( 
    {\displaystyle{\beta \hbar w \over 2}} \right) \right]  
  - {\displaystyle{\hbar  \over 2}} \sum_{i=1}^{2}a_{i}^{2}
    \Omega_{i}\coth \left( {\displaystyle{\beta \hbar \Omega _{i} 
    \over 2}} \right) 
\nonumber \\
&&- {\displaystyle{\sqrt{\pi }e^{2} \over D}}
  e^{-d^{2}/(2D)} \left[ I_{0}\left({\displaystyle{d^{2} \over 2D}} 
  \right) +I_{1}\left( {\displaystyle{d^{2} \over 2D}} \right) \right]  
\label{F} \\
&& -{\displaystyle{1 \over 2\pi \hbar \beta }}
   \int_{1/R_{b}}^{\infty }dq q|M_{q}|^{2}\int_{0}^{\hbar \beta /2}d\tau
   {\displaystyle{\cosh [\omega (q)(\tau -\hbar \beta /2)] \over 
   \sinh [\beta \hbar \omega (q)/2]}}
\nonumber \\
&&\times \exp \left[ -
   {\textstyle{\hbar q^{2} \over 2m_e}} \sum_{j=1}^{2}a_{j}^{2}
   {\textstyle{\cosh (\hbar \Omega _{j}\beta /2)-\cosh [\hbar 
   \Omega_{j}(\tau -\beta /2)] \over \Omega _{j}\sinh (\hbar 
   \Omega_{j}\beta/2)}}
\right]. \nonumber 
\end{eqnarray}
In this expression, $I_{0}$ and $I_{1}$ are Bessel functions of imaginary
argument, and 
\begin{equation}
D={\displaystyle{\hbar  \over m_e}} \sum_{j=1}^{2}
  {\displaystyle{a_{j}^{2} \over \Omega _{j}}}
  \coth \left( \hbar \Omega _{j}\beta /2\right),
\end{equation}
\begin{equation}
a_{1}=\sqrt{
      {\displaystyle{\Omega _{1}^{2}-w^{2} \over \Omega _{1}^{2}-
      \Omega_{2}^{2}}} };
a_{2}=\sqrt{
      {\displaystyle{w^{2}-\Omega _{2}^{2} \over \Omega _{1}^{2}-
      \Omega_{2}^{2}}}
}.
\end{equation}
Finally, $\Omega _{1}$ and $\Omega _{2}$ are the eigenfrequencies of the
trial system, given by 
\begin{equation}
  \Omega _{1,2}^{2}={\displaystyle{1 \over 2}}
  \left[ \Omega ^{2}+w^{2}\pm \sqrt{\left( \Omega ^{2}-w^{2}\right)
  ^{2}+4K/(Mm_e)}\right] .
\end{equation}
Optimal values of the variational parameters are determined by the
numerical minimization of the variational functional $F$ as given by
expression (\ref {F}). As the reader may notice, the result of the
variational pa\-th-in\-te\-gral me\-thod is slightly less intuitive than 
that of the strong-coupling approach of the previous section, nevertheless 
it is much more general and will allow us to introduce temperature to 
examine the melting of the Wigner lattice of ripplopolarons in the next 
section.

\section{Melting of the ripplopolaron Wigner lattice}
\label{sec:4}

The Lindemann melting criterion \cite{LindemanZPhys11} states in general
that a crystal lattice of objects (be it atoms, molecules, electrons, or
ripplopolarons) will melt when the average motion of the objects around
their lattice site is larger than a critical fraction $\delta _{0}$ of the
lattice parameter $d$. It would be a strenuous task to calculate from
first principles the exact value of the critical fraction $\delta _{0}$,
but for the particular case of electrons on a helium surface, we can make
use of an experimental determination. Grimes and Adams \cite{GrimesPRL42}
found that the Wigner lattice melts when $\Gamma =137\pm 15$, where
$\Gamma $ is the ratio of potential energy to the kinetic energy per
electron. In their experiment, the electron density varied from $10^{8}$
cm$^{-2}$ to $3\times 10^{8}$ cm$^{-2}$ while the melting temperature
$T_{c}$ varied from 0.23 K to 0.66 K. At temperature $T$ the average 
kinetic energy in a lattice potential $V_{lat}$ is
\begin{equation}
  E_{kin}={\displaystyle{\hbar \omega _{lat} \over 2}} \coth \left( 
  {\displaystyle{\hbar \omega _{lat} \over 2k_{B}T}} \right) ,
\end{equation}
and the average distance that an electron moves out of the lattice site is
determined by 
\begin{equation}
  \left\langle {\bf r}^{2}\right\rangle =
  {\displaystyle{\hbar  \over m_{e}\omega _{lat}}} \coth \left( 
  {\displaystyle{\hbar \omega _{lat} \over 2k_{B}T}}
  \right) = {\displaystyle{2E_{kin} \over m_{e}\omega _{lat}^{2}}} .
\end{equation}
From this we find that for the melting transition in Grimes and Adams' 
experiment \cite{GrimesPRL42}, the critical fraction equals $\delta 
_{0}\approx 0.13$. This estimate is in agreement with previous (empirical) 
estimates yielding $\delta _{0}\approx 0.1$ \cite{BedanovPRB49}, and we 
shall use it in the rest of this paper.

The unmodified Lindemann criterion as stated above cannot be applied to an
infinite layer of electrons on helium at non-zero temperature, because
(when a thermal occupation of the ripplon modes is assumed) a
straightforward calculation of the average distance that an electron moves
out of its lattice site yields a divergent result. This divergence is
closely related to Hohenberg's theorem forbidding Bose-Einstein
condensation in 2D. Therefore, many authors rely on a modified Lindemann
criterion \cite{BedanovPLA109} that considers the average distance between
two nearest neighbors instead of the average distance of a lattice
resident from its lattice site. However, for the current geometry this
modification is unnecessary: the multielectron bubble is a finite and
confined system, for which considerations based on Hohenberg's theorem do
not apply. Hence, we shall use the unmodified Lindemann criterion to study
the melting of the ripplopolaron lattice. In practice, we see that the
above-mentioned divergence is not present because there is a natural
cut-off wavelength for the ripplons: the lowest ripplon mode on a sphere
corresponds to an $\ell =1$ spherical harmonic, to which a characteristic
wavelength of the order of $1/R_{b}$ can be associated. We have checked
that the results do not depend on the precise value of the cut-off
wavelength $\lambda /R_{b}$ with $\lambda $ on the order of 1.

Within the approach of Fratini and Qu\'{e}merais \cite{FratiniEPJB14}, the
Wigner lattice of (ripplo)polarons melts when at least one of the two
following Lindemann criteria are met: 
\begin{equation}
\delta _{r}=
{\displaystyle{\sqrt{\left\langle {\bf R}_{cms}^{2}\right\rangle } \over d}}
>\delta _{0},  \label{Lind1}
\end{equation}
\begin{equation}
\delta _{\rho }=
{\displaystyle{\sqrt{\left\langle {\bf \rho }^{2}\right\rangle } \over d}}
>\delta _{0}.  \label{Lind2}
\end{equation}
where ${\bf \rho }$ and ${\bf R}_{cms}$ are, respectively, the
relative coordinate and the center of mass coordinate of the model system (%
\ref{L0}): if ${\bf r}$ is the electron coordinate and ${\bf R}$ is the
position coordinate of the fictitious ripplon mass $M$, this is 
\begin{equation}
{\bf R}_{cms}=
{\displaystyle{m_{e}{\bf r}+M{\bf R} \over m_{e}+M}}
;{\bf \rho }={\bf r}-{\bf R.}
\end{equation}
The appearance of two Lindemann criteria takes into account the composite
nature of (ripplo)polarons. As follows from the physical sense of the
coordinates ${\bf \rho }$ and ${\bf R}_{cms}$, the first criterion (%
\ref{Lind1}) is related to the melting of the ripplopolaron Wigner lattice
towards a ripplopolaron liquid, where the ripplopolarons move as a whole,
the electron together with its dimple. The second criterion (\ref{Lind2}) is
related to the dissociation of ripplopolarons: the electrons shed their
dimple.

The path-integral variational formalism outlined in the previous section
allows to calculate the expectation values $\left\langle 
{\bf R}_{cms}^{2}\right\rangle$ and 
$\left\langle {\bf \rho }^{2}\right\rangle $ with
respect to the ground state of the variationally optimal model system. We
find 
\begin{eqnarray}
\left\langle {\bf R}_{cms}^{2} \right\rangle &=&
 {\displaystyle{\hbar w^{4} \over m_e
 \left[ w^{2}(\Omega_{1}^{2}+\Omega _{2}^{2})
        -\Omega_{1}^{2}\Omega _{2}^{2} \right] 
 \left(\Omega_{1}^{2}-\Omega _{2}^{2}\right) }} 
\nonumber \\
 & & \times \left[ \Omega _{2}^{4}(\Omega _{1}^{2}-w^{2})\coth (\hbar 
     \Omega_{1}\beta /2)/\Omega_{1} \right.
\nonumber \\
 & & \left. + \Omega_{1}^{4} (w^{2}-\Omega_{2}^{2})
     \coth(\hbar \Omega_{2}\beta /2)/ \Omega_{2} \right]
,  \label{Rcms}
\end{eqnarray}
\begin{eqnarray}
\left\langle {\bf \rho }^{2}\right\rangle &=& 
{\displaystyle{\hbar  \over m_e
  \left( \Omega_{1}^{2}-\Omega_{2}^{2} \right) \left( \Omega_{1}^{2}
  -w^{2}\right) \left( w^{2}-\Omega_{2}^{2}\right) }}
\nonumber \\
  & & \times \left[ \Omega _{1}^{3}(w^{2}-\Omega _{2}^{2})\coth 
      \left(\hbar \Omega _{1}\beta /2\right) \right.
\nonumber \\
  & & \left. +\Omega _{2}^{3}(\Omega _{1}^{2}-w^{2})
      \coth(\hbar \Omega _{2}\beta /2) \right]
.  \label{rho}
\end{eqnarray}
The procedure to find whether the Lindeman criteria are fulfilled is then as
follows: first the optimal values of the variational parameters are obtained
by minimization of the free energy (\ref{F}), and then these optimal values
are substituted in (\ref{Rcms}),(\ref{rho}). Numerical calculation shows
that for ripplopolarons in an MEB the inequality $\Omega _{1}\gg w$ is
fulfilled ($w / \Omega _{1} \approx 10^{-3}$ to $10^{-2}$) so that the
strong-coupling regime is realized, in agreement with the results of Sec.
II. Owing to this inequality, we find from Eqs. (\ref{Rcms}),(\ref{rho})
that 
\begin{equation}
  \left\langle {\bf R}_{cms}^{2}\right\rangle \ll \left\langle {\bf 
  \rho }^{2}\right\rangle .
\end{equation}
So, the destruction of the ripplopolaron Wigner lattice in an MEB occurs
through the dissociation of ripplopolarons, since the second criterion (\ref
{Lind2}) will be fulfilled before the first (\ref{Lind1}). The results for
the melting of the ripplopolaron Wigner lattice are summarized in the phase
diagram shown in Fig. \ref{fig:3}. For every value of $N$, pressure $p$
and temperature $T$ in an experimentally accessible range, this figure shows
whether the ripplopolaron Wigner lattice is present (points above the
surface) or molten (points below the surface). Below a critical pressure (on
the order of 10$^{4}$ Pa) the ripplopolaron solid will melt into an electron
liquid. This critical pressure is nearly independent of the number of
electrons (except for the smallest bubbles) and is weakly temperature
dependent, up to the helium critical temperature 5.2 K$.$ This can be
understood since the typical lattice potential well in which the
ripplopolaron resides has frequencies of the order of THz or larger, which
correspond to $\sim 10$ K.

\section{Discussion}
\label{sec:5}

In the previous section we have established that the ripplopolaron Wigner
lattice will not melt into a liquid of ripplopolarons, but rather melt
through dissociation of the composite quasiparticle that is the
ripplopolaron. The absence of a ripplopolaron liquid phase can be understood
intuitively from the fact that the ripplon frequencies (typically GHz) are
several orders of magnitude smaller than the electron frequencies in the
lattice potential (typically THz). In order to create a liquid of
ripplopolarons, the ripplopolarons have to move with an average velocity
large enough to keep the ripplopolaron lattice molten. This motion has to be
of the entire object, namely the electron and its dimple. But, at the
velocities required to keep the ripplopolaron liquid from freezing into a
lattice, the dimples cannot follow the electrons. Thus, ripplopolarons only
exist in a crystallized state.

The present treatment does not allow us to derive the stucture of this
lattice -- the mean-field approximation made for the lattice potential
prohibits this. The problem of the exact lattice structure is complicated
by the topology of the surface \cite{PerezPRB56}: unlike for a flat
surface, it is impossible to tile a sphere with a triangular lattice;
frustration of the lattice in the form of point defects is unavoidable,
providing nucleation points for melting the lattice. The problem of
placing classical point charges on a sphere was first considered by 
Thomson \cite{ThomsonPhilMag7} and was recently reconsidered for localized
electrons in multielectron bubbles \cite{LenzPRL87}.

The present treatment does allow us to study also the electron Wigner
lattice, by putting $e|{\bf E|}=0$ in the above results, thus switching
the electron-ripplon coupling off. A Wigner lattice of electrons is to be
distinguished from a Wigner lattice of ripplopolarons. The lattice of
ripplopolarons on the one hand melts through dissociation of the
ripplopolarons, and this melting line is almost temperature independent.
The lattice of electrons on the other hand melts though either classical
thermal motion (when the temperature reaches a melting temperature of
about 0.5 K), or through quantum melting when the density of electrons is
large enough so that the extent of the zero-point motion becomes
comparable to the lattice parameter. In the Wigner lattice of the
ripplopolarons on the one hand, the particles are localized by the
self-induced polaronic trapping potential (the dimple) due to the
electron-ripplon interaction. In the Wigner lattice of electrons, the
electrons are localized through the Coulomb interaction between the
electrons. Finally, the region in phase space where the ripplopolaron
Wigner lattice resides is different from the region where the electron
Wigner lattice is found. This is illustrated in Fig. \ref{fig:4}, where
the phase diagram drawn in Fig. \ref{fig:3} is extended to huge bubbles
(approaching the flat surface geometry). In the corner of largest $N$
($N>10^{9},$ $R_{b}\gtrsim 1$ mm, $n_s \lesssim 10^9-10^{10}$ cm$^{-2}$)
and lowest pressure ($p<0.1$ Pa), we find that an electron Wigner lattice
(without individual dimples) can still be formed below $T=0.4$ K. Thus,
the electron Wigner lattice is recovered and the melting temperature
derived from our treatment is in agreement with the experimentally
observed temperature \cite{GrimesPRL42}. Our calculations show that, as
the bubble is compressed, the electron Wigner lattice will quantum melt
because of the increased density of electrons. The region of phase space
where the electron Wigner lattice is present is separated from the region
where the ripplopolaron Wigner lattice is present by a region where the
predicted phase is an electron liquid. The ripplopolaron liquid phase, as
mentioned before, does not exist. Since, as mentioned in the previous
paragraph, our method does not allow us to study the crystal structure, we
cannot, in the electron Wigner lattice phase, distinguish between the
crystalline and the hexatic phase \cite{NelsonPRB19}.

The new phase that we predict, the ripplopolaron Wigner lattice, will not
be present for electrons on a flat helium surface. At the values of the
pressing field necessary to obtain a strong enough electron-ripplon
coupling, the flat helium surface is no longer stable against
long-wavelength deformations \cite{GorkovJETP18}. Multielectron bubbles,
with their different ripplon dispersion and the presence of stabilizing
factors such as the energy barrier against fissioning \cite{TemperePRB67},
allow for much larger electric fields pressing the electrons against the
helium surface. The regime of $N$,$p$,$T$ parameters suitable for the
creation of a ripplopolaron Wigner lattice lies within the regime that would
be achievable in recently proposed experiments aimed at stabilizing
multielectron bubbles \cite{SilveraProc}. The ripplopolaron Wigner lattice
and its melting transition might be detected by spectroscopic techniques 
\cite{GrimesPRL42,FisherPRL42} probing for example the transverse phonon
modes of the lattice \cite{DevillePRL53}. 

\section{Conclusions}

In this paper, we investigate the properties of ripplopolarons in a
multielectron bubble in helium using path-integral methods similar to those
developed for a lattice of polarons \cite{FratiniEPJB14}. Expressions are
derived for the free energy and the dimple shape of ripplopolarons in a
Wigner lattice in a multielectron bubble, as a function of temperature,
externally applied pressure and number of electrons in the bubble. We find
that, owing to the difference in the ripplon and longitudinal plasmon
frequencies \cite{KliminSSC}, the ripplopolarons exist only in a Wigner
crystallized state. This state differs from the Wigner lattice of electrons,
in that the electrons in the ripplopolaron Wigner lattice are localized by
the electron-ripplon interaction rather than the Coulomb repulsion, and in
that the melting occurs through the dissociation of the ripplopolarons. As
electron-ripplon interaction is weakened (for example by reducing the
externally applied pressure on the multielectron bubble) the electrons can
shed their localized dimple and the ripplopolaron Wigner state is destroyed.
The melting transition is shown to occur in a region of phase space that is
accessible to recently proposed experiments for stabilizing multielectron
bubbles.

{\bf Acknowledgements} - The authors would like to acknowlegde fruitful and intensive discussions with V.M. Fomin. Discussions with J. Huang are gratefully
acknowledged. J. T. is supported financially by the FWO-Flanders. This
research has been supported by the Department of Energy, Grant
DE-FG02-ER45978, and by the GOA BOF UA 2000, IUAP, the FWO-V projects Nos.
G.0435.03, G.0274.01, G.0306.00.

\bigskip 

\appendix

\section{Electron-ripplon interaction in MEBs}

\subsection{Ripplon dispersion on a sphere}

The modes of oscillation of the bubble surface will be called `ripplons'
in analogy with the surface oscillations on a flat helium surface. In
general, the deformed bubble surface can be described by a function
$R(\Omega )$ which gives the distance, from the center of the bubble, of
the bubble surface in the direction determined by the spherical angles
$\Omega =\{\theta ,\phi \}$. The deformation $u(\Omega )$ from spherical
symmetry can be expanded in spherical harmonics $Y_{\ell m}(\Omega )$:  
\begin{equation} 
  R(\Omega )=R_{b}+u(\Omega )=R_{b}+\sum_{\ell ,m}
  Q_{\ell,m} Y_{\ell ,m}(\Omega ). \label{radius} 
\end{equation} 
In this expression, $R_{b}$ is the angle-averaged radius, and $Q_{\ell m}$ 
is the amplitude of the deformational mode corresponding to the spherical
harmonic $Y_{\ell ,m}$. In the summation, we abbreviate 
$\sum_{\ell,m}=\sum_{\ell =1}^{\infty }\sum_{m=-\ell }^{\ell }$. 
The equilibrium bubble radius is found by balancing the surface tension 
and pressure terms with the Coulomb repulsion: $R_{b}$ satisfies 
\begin{equation} 
  2\sigma R_{b}+pR_{b}^{2}= \frac{e^2}{4\pi\varepsilon R_{b}^{2}},
\end{equation} 
with $\sigma \approx 3.6\times
10^{-4}$ J/m$^{2}$ the surface tension of helium, $\varepsilon =1.0572$
the dielectric constant of helium, $p$ the difference in pressure outside
and inside the bubble, $e$ the electron charge and $N$ the number of
electrons in the bubble. Expanding the energy of the bubble up to second
order in $Q_{\ell m}$ allows to derive the frequency of oscillation of a
particular mode of deformation \cite {TemperePRL87}:  
\begin{equation}
 \omega (\ell) = \sqrt{ 
  \frac{\ell +1}{\rho R_{b}^{3}}
  \left[ \sigma (\ell ^{2}+\ell +1)+pR_{b}-
   {\textstyle{N^{2}e^{2} \over 4\pi \varepsilon R_{b}^{3}}}
   {\textstyle{\ell^{2}-\varepsilon (\ell +1) \over 
   \ell +\varepsilon (\ell +1)}}
  \right] }, \label{ripfreq} 
\end{equation} 
where $\rho \approx 145$ kg/m$^{3}$ the mass density of helium. Thus, for
small amplitude deformations of the bubble, we find that the shape of the
bubble oscillates with frequencies given by (\ref{ripfreq}). Taking the
bare ripplon frequency (without the effect of the interaction with
electrons) and putting $R_{b}\rightarrow \infty $ (with $\ell /R_{b}=q$ a
constant) we find
\begin{equation}
 \omega _{R_{b}\rightarrow \infty }^{bare}(q)=
 \sqrt{ \frac{\sigma}{\rho} q^3+ \frac{pR_{b}}{\rho} q}.
\end{equation}
This dispersion relation corresponds to the ripplon dispersion on the flat
surface, with the difference that there is in our dispersion relation no
gravitational term, but a term related to the pressure on the bubble. 

\subsection{Electron ripplon interaction in the MEB}

The interaction energy between the ripplons and the electrons in the
multielectron bubble can be derived from the following considerations: (i)
the distance between the layer electrons and the helium surface is fixed
(the electrons find themselves confined to an effectively 2D 
surface anchored to the helium surface \cite{SalomaaPRL47}) and (ii) the
electrons are subjected to a force field, arising fromt the electric field
of the other electrons. For a spherical bubble, this electric field lies
along the radial direction and equals 
\begin{equation}
  {\bf E}=-{\frac{Ne}{2R_{b}^{2}}} {\bf e}_{{\bf r}}.
\end{equation}
A bubble shape oscillation will displace the layer of electrons anchored to
the surface. The interaction energy which arises from this, equals the
displacement of the electrons times the force $e{\bf E}$ acting on them.
Thus, we get for the interaction Hamiltonian 
\begin{equation}
\hat{H}_{int}=\sum_{j}e|{\bf E}|\times u(\hat{\Omega}_{j}).
\end{equation}
Here $u(\Omega )$ is the radial displacement of the surface in the direction
given by the spherical angle $\Omega $; and $\hat{\Omega}_{j}$ is the
(angular) position operator for electron $j$. The displacement can be
rewritten using (\ref{radius}) and we find 
\begin{equation}
\hat{H}_{int}=\sum_{j}e|{\bf E}|\sum_{\ell ,m}\hat{Q}_{\ell m}Y_{\ell
m}(\hat{\Omega}_{j}).
\end{equation}
Using the relation 
\begin{equation}
  \hat{Q}_{\ell ,m}=(-1)^{(m-|m|)/2} \sqrt{
  {\textstyle{\hbar (\ell +1) \over 2\rho R_{b}^{3}\omega _{\ell }}}} 
  (\hat{a}_{\ell ,m}+\hat{a}_{\ell ,-m}^{+}), \label{Qriplon}
\end{equation}
the interaction Hamiltonian can be written in the suggestive form 
\begin{equation}
\hat{H}_{int}=\sum_{\ell ,m}\sum_{j}M_{\ell ,m}Y_{\ell ,m}(\hat{\Omega%
}_{j})(\hat{a}_{\ell ,m}+\hat{a}_{\ell ,-m}^{+}),  \label{Hint}
\end{equation}
with the electron-ripplon coupling amplitude for a MEB given by 
\begin{equation}
  M_{\ell ,m}=(-1)^{(m-|m|)/2} {\displaystyle{Ne^{2} \over 2R_{b}^{2}}}
  \sqrt{{\displaystyle{\hbar (\ell +1) \over 2\rho R_{b}^{3}
  \omega _{\ell }}} }
\end{equation}

\subsection{Locally flat approximation}

Substituting $M_{\ell ,m}$ into (\ref{Hint}), we get 
\begin{eqnarray}
\lefteqn{ \hat{H}_{int}=\sum_{\ell ,m}\sum_{j}
  {\frac{Ne^{2} }{ 2R_{b}^{2}}}
\sqrt{
  {\frac{\hbar (\ell +1) }{ 2\rho R_{b}^{3}\omega _{\ell }}}
  } 
} & & \\
  & & \times \left[ (-1)^{(m-|m|)/2}
  {\frac{Y_{\ell ,m}(\hat{\Omega}_{j}) }{ R_{b}}}
  \right] (\hat{a}_{\ell ,m}+\hat{a}_{\ell ,-m}^{+}) 
. \nonumber
\end{eqnarray}
In this expression, we consider the limit of a bubble so large that the
surface becomes flat on all length scales of interest. Hence we let $%
R_{b}\rightarrow \infty $ but keep $\ell /R_{b}=q$ a constant. This means we
have to let $\ell \rightarrow \infty $ as well. In this limit, 
\begin{equation}
  \lim_{\ell \rightarrow \infty }Y_{\ell ,0}(\theta )=%
  {\displaystyle{i^{\ell } \over \pi \sqrt{\sin \theta }}}%
  \sin [(\ell +1/2)\theta +\pi /4],
\end{equation}
and $Y_{\ell ,0}(\theta )$ varies locally as a plane wave with wave vector $%
q=\ell /R_{b}$. The wave function $Y_{\ell ,m}(\hat{\Omega}_{j})/R_{b}$ is
furthermore normalised with respect to integration over the surface (with
total area $4\pi R_{b}^{2}$). Thus, we get in the locally flat approximation 
\begin{equation}
  \hat{H}_{int}=\sum_{{\bf q}}\sum_{j}
  {\displaystyle{Ne^{2} \over 2R_{b}^{2}}} \sqrt{
  {\displaystyle{\hbar q \over 2\rho \omega (q)}}%
  }e^{i{\bf q}.{\bf \hat{r}}_{j}}(\hat{a}_{{\bf q}}
  +\hat{a}_{-{\bf q}}^{+}),
\end{equation}
or 
\begin{eqnarray}
  \hat{H}_{int} &=&\sum_{{\bf q}}\sum_{j}M_{q}e^{i{\bf q}.{\bf \hat{r}}%
  _{j}}(\hat{a}_{{\bf q}}+\hat{a}_{-{\bf q}}^{+}),  
\nonumber \\
  M_{q} &=&e|{\bf E|}\sqrt{%
  {\displaystyle{\hbar q \over 2\rho \omega (q)}} }.
\end{eqnarray}
This corresponds in the limit of large bubbles to the interaction
Hamiltonian expected for a flat surface.

\begin{table}
 \caption{
   For typical multielectron bubbles, the values of
   the several physical quantities are given at zero applied external  
   pressure. The number of electrons in the bubble ($N$), the bubble 
   radius ($R_{b}$), the average interelectron distance ($d$), 
   the surface density of electrons ($n_{s}$), the pressing field 
   generated by the electrons $|{\bf E|}$ (see Eq. \ref{pressfield}), the 
   characteristic energy scale of the electron-ripplon interaction 
   [$(e|{\bf E}|)^{2}/\sigma $, cf. formula (\ref{Eripol})], and the 
   characteristic frequency of the lattice potental ($\omega _{%
   lat}$) are given. Compare these quantities to, for example, the 
   maximum density ($\approx 2\times 10^{9}$ cm$^{-2}$) and the maximum 
   pressing field ($\approx 3$ kV/cm) achievable on a flat helium 
   surface over bulk. }
 \label{tab:1}       
 \begin{tabular}{cccc}
    $N$ & $R_{b}$($\mu $m) & $d$(nm) & $n_{s}$(cm$^{-2}$) 
 \\ \hline
    10$^{3}$ & 0.228 & 25.57 & 1.529$\times 10^{11}$ 
 \\
    10$^{5}$ & 4.937 & 55.34 & 3.265$\times 10^{10}$ 
 \\
    10$^{7}$ & 106.4 & 119.3 & 7.025$\times 10^{9}$ 
 \\ \hline\hline
    $N$ & $|{\bf E|}$(kV/cm) & $e^{2}|{\bf E}|^{2}/\sigma $ (meV) & 
     $\omega_{lat}$(THz) 
 \\ \hline
    10$^{3}$ & 138.3 & 85.16 & 3.891 
 \\
   10$^{5}$ & 63.80 & 3.884 & 1.222 
 \\
   10$^{7}$ & 6.350 & 0.180 & 0.386
\end{tabular}
\end{table}

\begin{figure}[tbp]
  \caption{ The variational parameter $a$ describing the width of
    the electron wave function in the strong-coupling approach as a 
    function of number of electrons and pressure in the multielectron 
    bubble. $d$ is the interelectron separation. }
  \label{fig:1}  
\end{figure}

\begin{figure}[tbp]
  \caption{ For a bubble with $N=10^5$ electrons, at different
    pressures the shape of the dimpled surface is shown. Electrons are 
    present on the surface, separated from eachother by the lattice 
    parameter $d$. Underneath each electron there is an individual dimple, 
    induced by the electron-ripplon interaction. As the pressure is 
    increased, the bubble radius decreases, and the electron-ripplon 
    interaction becomes stronger, resulting in a stronger dimpling 
    effect. }
  \label{fig:2}  
\end{figure}

\begin{figure}[tbp]
  \caption{ The phase diagram for the spherical 2D layer of
     electrons in the MEB. Above a critical pressure, a ripplopolaron 
     solid (a Wigner lattice of electrons with dimples in the helium 
     surface underneath them) is formed. Below the critical pressure, the 
     ripplopolaron solid melts into an electron liquid through 
     dissociation of ripplopolarons. }
  \label{fig:3}
\end{figure}

\begin{figure}[tbp]
  \caption{ The phase diagram shown in Fig. 3 is extended to reveal
     the relation of the ripplopolaron Wigner lattice to the Wigner 
     lattice of electrons. These are distinct, not only in melting 
     properties (the ripplopolaron Wigner lattice melts through 
     dissociation of ripplopolarons), but also in their location on the 
     phase diagram.  The region for the Wigner lattice of electrons 
     without dimples --in agreement with the observation of Grimes and 
     Adams [22]-- starts at large $N$ and is quantum 
     molten by pressurizing the bubble. }
  \label{fig:4}
\end{figure}   

\end{document}